%% file: SMP-12-027_temp.tex
\pdfoutput=1

\documentclass[11pt,twoside,a4paper,cmspaper,final,collab]{cms-tdr}
\def\svnVersion{287184}\def\cmsCernNo{2013-037}\def\cmsCernDate{\today}\def\cmsMessage{Published in the European Physical Journal C as \href{http://dx.doi.org/10.1140/epjc/s10052-015-3376-y}{\doi{10.1140/epjc/s10052-015-3376-y.}}}

\begin{document}\cmsNoteHeader{SMP-12-027}

\hyphenation{had-ron-i-za-tion}
\hyphenation{cal-or-i-me-ter}
\hyphenation{de-vices}
\RCS$Revision: 281193 $
\RCS$HeadURL: svn+ssh://svn.cern.ch/reps/tdr2/papers/SMP-12-027/trunk/SMP-12-027.tex $
\RCS$Id: SMP-12-027.tex 281193 2015-03-18 19:55:06Z alverson $
\newlength\cmsFigWidth
\ifthenelse{\boolean{cms@external}}{\setlength\cmsFigWidth{0.98\columnwidth}}{\setlength\cmsFigWidth{0.6\textwidth}}
\ifthenelse{\boolean{cms@external}}{\providecommand{\cmsLeft}{top\xspace}}{\providecommand{\cmsLeft}{left\xspace}}
\ifthenelse{\boolean{cms@external}}{\providecommand{\cmsRight}{bottom\xspace}}{\providecommand{\cmsRight}{right\xspace}}
\ifthenelse{\boolean{cms@external}}{\providecommand{\cmsTopLeft}{top\xspace}}{\providecommand{\cmsTopLeft}{top left\xspace}}
\ifthenelse{\boolean{cms@external}}{\providecommand{\cmsTopRight}{centre\xspace}}{\providecommand{\cmsTopRight}{top right\xspace}}

\providecommand{\alps}{\ensuremath{\alpha_\mathrm{S}}\xspace}
\providecommand{\alpsmz}{\ensuremath{\alpha_S(M_\cPZ)}\xspace}
\providecommand{\alpsq}{\ensuremath{\alpha_\mathrm{S}(Q)}\xspace}
\providecommand{\chisq}{\ensuremath{\chi^2}\xspace}
\providecommand{\chisqndof}{\ensuremath{\chi^2/n_\mathrm{dof}}\xspace}
\providecommand{\HOPPET} {{\textsc{hoppet}}\xspace}
\providecommand{\NLOJETPP} {{\textsc{NLOJet++}}\xspace}
\providecommand{\fastNLO} {{\textsc{fastNLO}}\xspace}
\providecommand{\fastjet} {{\textsc{FastJet}}\xspace}
\providecommand{\PYTHIAE} {{\textsc{pythia8}}\xspace}
\providecommand{\RunDec} {{\textsc{RunDec}}\xspace}
\providecommand{\mjjj}{\ensuremath{m_3}\xspace}
\providecommand{\ymax}{\ensuremath{y_{\text{max}}}\xspace}
\providecommand{\aymax}{\ensuremath{\abs{ y}_{\text{max}}}\xspace}
\providecommand{\Ratio}{\ensuremath{R_\mathrm{32}}\xspace}
\providecommand{\mur}{\ensuremath{\mu_r}\xspace}
\providecommand{\muf}{\ensuremath{\mu_f}\xspace}
\providecommand{\texp}{\,\text{(exp)}\xspace}
\providecommand{\scale}{\,\text{(scale)}\xspace}

\providecommand{\rbthm}{\rule[-2ex]{0ex}{5ex}}
\providecommand{\rbtrr}{\rule[-0.8ex]{0ex}{3.2ex}}
\cmsNoteHeader{SMP-12-027}
\title{Measurement of the inclusive 3-jet production differential cross section in proton-proton
  collisions at 7\TeV and determination of the strong coupling constant in the \TeV range}
\titlerunning{3-jet differential cross section in pp at 7\TeV and determination of \alps}
\address[ekp]{Institut f\"ur Experimentelle Kernphysik, Karlsruhe
  Institute of Technology}

\date{\today}

\abstract{This paper presents a measurement of the inclusive
  3-jet production differential cross section at a proton-proton centre-of-mass energy of
  7\TeV using data corresponding to an integrated
  luminosity of 5\fbinv collected with the CMS detector.
  The analysis is based on the three jets with the highest transverse momenta.
  The cross section is measured as a function of the
  invariant mass of the three jets in a range of 445--3270\GeV and in two bins
  of the maximum rapidity of the jets up to a value of 2.
  A comparison between the measurement and the prediction from
  perturbative QCD at next-to-leading order is performed. Within
  uncertainties, data and theory are in agreement. The sensitivity of
  the observable to the strong coupling constant \alps is studied.
  A fit to all data points with 3-jet
  masses larger than 664\GeV gives a value of the strong coupling
  constant of
  $\alpsmz = 0.1171
  \allowbreak\pm 0.0013\texp
  \allowbreak\,^{+0.0073}_{-0.0047}\thy$.}

\hypersetup{
pdfauthor={CMS Collaboration},
pdftitle={Measurement of the inclusive 3-jet production differential cross section in proton-proton
  collisions at 7 TeV and determination of the strong coupling constant in the TeV range},
pdfsubject={CMS},
pdfkeywords={CMS, physics, QCD, jets, 3-jet mass, PDF, strong coupling constant, alpha-S}}

\maketitle
\section{Introduction}
\label{sec:intro}

A key characteristic of highly energetic proton-proton collisions at
the LHC is the abundant production of
multijet events. At high transverse momenta \pt, such events are
described by quantum chromodynamics (QCD) in terms of parton-parton
scattering. The simplest jet production process corresponds to a $2
\to 2$ reaction with the two outgoing partons fragmenting into
a pair of jets. Two cross sections, for which the leading-order (LO)
predictions in perturbative QCD (pQCD) are proportional to the square
of the strong coupling constant, $\alps^2$, are conventionally
defined: the inclusive single-jet cross section as a function of jet \pt and
rapidity $y$, and the 2-jet production cross section as a function of
the 2-jet invariant mass and a rapidity-related kinematic quantity
that provides a separation of the phase space into exclusive bins. The
ATLAS Collaboration usually characterizes the 2-jet system in terms of
the rapidity separation of the two jets leading in \pt, while CMS
employs the larger of the two absolute rapidities of the two jets.
Corresponding measurements by the ATLAS and CMS Collaborations can be
found in Refs.~\cite{Aad:2013tea, Aad:2013lpa, CMS-PAPERS-QCD-11-004,
  CMS-PAPERS-QCD-10-011, CMS-PAPERS-QCD-10-025, Aad:2010wv}.

In this paper, the inclusive 3-jet production differential cross section is
measured as a function of the invariant mass \mjjj of the three jets
leading in \pt and of their maximum rapidity \ymax, which are defined
as follows:
\ifthenelse{\boolean{cms@external}}{
\begin{equation}\begin{split}
  \label{eq:mjjj}
  \mjjj^{2} =&  \left( p_{1}+p_{2}+p_{3}\right)^{2}\\
  \ymax     =&  \sgn\big(
  \abs{\max(y_1, y_2, y_3)} \\
  &-
  \abs{\min(y_1, y_2, y_3)}\big) \cdot
  \max \left( \abs{y_1}, \abs{y_2}, \abs{y_3} \right),
\end{split}\end{equation}
}{
\begin{equation}\begin{split}
  \label{eq:mjjj}
  \mjjj^{2} & =  \left( p_{1}+p_{2}+p_{3}\right)^{2}\\
  \ymax     & =  \sgn\big(
  \abs{\max(y_1, y_2, y_3)} -
  \abs{\min(y_1, y_2, y_3)}\big) \cdot
  \max \left( \abs{y_1}, \abs{y_2}, \abs{y_3} \right),
\end{split}\end{equation}
}
where $p_i$ and $y_i$ are the four-momentum and rapidity of the $i$th
jet leading in \pt. Following Ref.~\cite{CMS-PAPERS-QCD-11-004},
\ymax is defined as a signed quantity such that the
double-differential cross section, $\rd^2\sigma/\rd{\mjjj}\,\rd{\ymax}$, can
be written in a way similar to the inclusive jet cross section,
$\rd^2\sigma/\rd{\pt}\,\rd{y}$, including a factor of 2 for rapidity bin
widths in terms of $\abs{\ymax}$ and $\abs{y}$, respectively.
The absolute value of \ymax is equal to the maximum $\abs{y}$ of the jets,
denoted \aymax. A previous study of the 3-jet mass spectra was published by the
D0~Collaboration~\cite{Abazov:2011ub}. Very recently, ATLAS submitted
a 3-jet cross section measurement~\cite{Aad:2014rma}.

For this cross section, the LO process is proportional to $\alps^3$ and
theoretical predictions are available up to next-to-leading order
(NLO)~\cite{Nagy:2001fj,Nagy:2003tz} making precise comparisons to
data possible. The potential impact of this measurement on the parton distribution
functions (PDFs) of the proton is studied and the strong coupling
constant \alps is extracted. In previous publications by CMS, the value
of \alps was determined to %
$\alpsmz = 0.1148 %
\allowbreak\pm 0.0014\texp %
\allowbreak\pm 0.0050\thy$ %
by investigating the ratio of inclusive 3-jet to inclusive 2-jet
production, \Ratio~\cite{CMS-PAPERS-QCD-11-003}, and %
$\alpsmz = 0.1185 %
\allowbreak\pm 0.0019\texp %
\allowbreak\,^{+0.0060}_{-0.0037}\thy$ %
by fitting the inclusive jet cross
section~\cite{CMS-PAPER-SMP-12-028}. %
The ratio \Ratio benefits from uncertainty cancellations, but it is only
proportional to $\alps$ at LO, leading to a correspondingly
high sensitivity to its experimental uncertainties in fits of
\alpsmz. The second observable, which is similar to the denominator
in \Ratio, is proportional to $\alps^2$ at LO with a
sensitivity to experimental uncertainties reduced by a factor of $1/2$, but
without uncertainty cancellations. It is interesting to study how fits
of \alps to the inclusive 3-jet mass cross section,
$\rd^2\sigma/\rd{\mjjj}\,\rd{\ymax}$, which is a 3-jet observable similar to
the numerator of \Ratio, compare to previous results.

The data analyzed in the following were recorded by the CMS detector
at the LHC during the 2011 data-taking period
at a proton-proton centre-of-mass energy of 7\TeV and correspond to an integrated
luminosity of 5.0\fbinv. Jets are clustered by using the infrared- and
collinear-safe anti-\kt algorithm~\cite{Cacciari:2008gp} as
implemented in the \fastjet package~\cite{Cacciari:2011ma} with a jet
size parameter of $R=0.7$. A smaller jet size parameter of $R=0.5$
has been investigated, but was found to describe the data less well.
Similarly, in Ref.~\cite{Chatrchyan:2014gia} it is shown that the inclusive
jet cross section is better described by NLO theory for $R=0.7$ than
for $R=0.5$.

Events are studied in which at least three jets are found up to a
rapidity of $\abs{y}=3$ that are above a minimal \pt threshold of 100\GeV.
The jet yields are corrected for detector effects resulting
in a final measurement phase space of $445\GeV \leq \mjjj < 3270\GeV$
and $\aymax < 2$.
Extension of the analysis to larger values of \aymax was not feasible
with the available trigger paths.

This paper is divided into seven parts. Section~\ref{sec:detector}
presents an overview of the CMS detector and the event reconstruction.
Sections~\ref{sec:selection} and~\ref{sec:measurement} discuss the
event selection and present the measurement. Theoretical ingredients
are introduced in Section~\ref{sec:theory} and are applied in
Section~\ref{sec:results} to determine \alpsmz from a fit to the
measured 3-jet production cross section. Conclusions are presented in
Section~\ref{sec:conclusions}.
\section{Apparatus and event reconstruction}
\label{sec:detector}

The central feature of the CMS apparatus is a superconducting solenoid
of 6\unit{m} internal diameter, providing a magnetic field of
3.8\unit{T}. Within the superconducting solenoid volume are a silicon
pixel and strip tracker, a lead tungstate crystal electromagnetic
calorimeter (ECAL), and a brass and scintillator hadron calorimeter
(HCAL), each composed of a barrel and two endcap sections. Muons are
measured in gas-ionization detectors embedded in the steel flux-return
yoke outside the solenoid. Extensive forward calorimetry complements
the coverage provided by the barrel and endcap detectors.

The first level (L1) of the CMS trigger system, composed of custom
hardware processors, uses information from the calorimeters and muon
detectors to select the most interesting events in a fixed time
interval of less than 4\mus. The high level trigger (HLT) processor
farm further decreases the event rate from around 100\unit{kHz} to
around 400\unit{Hz}, before data storage.

The particle-flow algorithm reconstructs and identifies each particle
candidate with an optimized combination of all subdetector
information~\cite{CMS-PAS-PFT-09-001,CMS-PAS-PFT-10-001}.
For each event, the reconstructed particle candidates are clustered
into hadronic jets by using the anti-\kt algorithm with a jet size
parameter of $R=0.7$. The jet momentum is determined as the vectorial
sum of all constituent momenta in this jet, and is found in the
simulation to be within 5\% to 10\% of the true momentum over the
whole \pt spectrum and detector acceptance.  An offset correction is
applied to take into account the extra energy clustered into jets due
to additional proton-proton interactions within the same or
neighbouring bunch crossings (pileup). Jet energy corrections are
derived from the simulation, and are confirmed with in situ
measurements with the energy balance of dijet, photon+jet, and
\cPZ+jet events~\cite{CMS-PAS-JME-10-010,CMS-PAPERS-JME-10-011}. The jet
energy resolution amounts typically to 15\% at 10\GeV, 8\% at 100\GeV,
and 4\% at 1\TeV. A more detailed description of the CMS apparatus can
be found in Ref.~\cite{Chatrchyan:2008aa}.
\section{Event selection}
\label{sec:selection}

The data set used for this analysis contains all events that were
triggered by any of the single-jet triggers. A single-jet trigger
accepts events if at least one reconstructed jet surpasses a
transverse momentum threshold. During the 2011 data-taking
period, triggers with eight different thresholds ranging from 60\GeV
to 370\GeV were employed. They are listed in
Table~\ref{tab:trigger} with the number of events recorded by each trigger
and the corresponding turn-on threshold
$p_\mathrm{T,99\%}$, where the trigger is more than 99\%
efficient.

\begin{table}[tbp]
  \topcaption{Trigger and turn-on thresholds in leading jet \pt,
    and the number of events recorded via the single-jet trigger
    paths used for this measurement.}
  \label{tab:trigger}
  \centering
  \begin{tabular}{cccc}
    Trigger threshold & Turn-on threshold & Recorded events\\
    \multicolumn{1}{c}{\pt [{\GeVns}]} &
    \multicolumn{1}{c}{$p_\mathrm{T,99\%}$ [\GeVns{}]} & \\\hline
    $ 60$  &  $85$ &  2\,591\,154 \\
    $ 80$  & $110$ &  1\,491\,011 \\
    $110$  & $144$ &  2\,574\,451 \\
    $150$  & $192$ &  2\,572\,083 \\
    $190$  & $238$ &  3\,533\,874 \\
    $240$  & $294$ &  3\,629\,577 \\
    $300$  & $355$ &  9\,785\,529 \\
    $370$  & $435$ &  3\,129\,458
  \end{tabular}
\end{table}

The different triggers are used to measure the 3-jet mass spectrum
in mutually exclusive regions of the phase space,
defined in terms of the \pt of the leading jet:
the \pt interval covered by a single-jet trigger starts at the
corresponding turn-on threshold $p_\mathrm{T,99\%}$ and ends at the
turn-on threshold of the trigger with the next highest threshold.
The final 3-jet mass spectrum is obtained by summing the
spectra measured with the different triggers
while taking trigger prescale factors into account.
Apart from the prescaling, the trigger efficiency is more than 99\%
across the entire mass range studied.

In the inner rapidity region, most single-jet triggers
contribute up to 50\% of the final event yield, with
the exception of the two triggers with the lowest and highest
threshold, which contribute up to 80\% and 100\% respectively, depending on $\mjjj$.
In particular, starting at 1100\GeV, the majority of
the events are taken from the highest unprescaled trigger.
In the outer rapidity region, each jet trigger contributes over a large range of
three-jet masses to the measurement. With the exception of the two triggers with the
lowest and highest thresholds, each trigger contributes around 25\% to the final event yield.

The recorded events are filtered with tracking-based
selections~\cite{CMS-PAS-TRK-10-005} to remove interactions between the
circulating proton bunches and residual gas particles or the beam
collimators. To further reject beam backgrounds and off-centre
parasitic bunch crossings, standard vertex selection cuts are
applied~\cite{CMS-PAS-TRK-10-005}. To enhance the QCD event purity,
events in which the missing transverse energy $\ETmiss$ amounts to more
than 30\% of the measured total transverse energy are removed. The
missing transverse energy is calculated by requiring momentum
conservation for the reconstructed particle flow
candidates~\cite{CMS-PAPERS-JME-10-011}.

Jet identification (jet ID) selection
criteria~\cite{CMS-PAS-JME-09-008} are developed to reject pure noise
or noise enhanced jets, while keeping more than 99\% of physical jets
with transverse momentum above 10\GeV.  In contrast to the previous
selection criteria, which reject complete events, the jet ID
removes only individual jets from the event.  The jet ID applied to the
particle-flow jets requires that each jet should contain at least two
particles, one of which is a charged hadron. In addition, the jet
energy fraction carried by neutral hadrons and photons must be less
than 90\%. These criteria have an efficiency greater than 99\% for
hadronic jets.
\section{Measurement and experimental uncertainties}
\label{sec:measurement}

The double-differential 3-jet production cross section is measured as
a function of the invariant 3-jet mass \mjjj and the maximum rapidity
\ymax of the three jets with the highest transverse momenta in the
event:

\begin{equation}
  \frac{\rd^{2}\sigma }{\rd{\mjjj}\,\rd\ymax} =
  \frac{1}{\epsilon \mathcal{L}} \frac{N}{\Delta \mjjj
    (2\Delta\aymax)}.
\end{equation}

Here, $\mathcal{L}$ is the integrated luminosity and $N$ is the number
of events. The efficiency $\epsilon$ is the product of the trigger and
event selection efficiencies, and differs from unity by less than one
percent for this jet analysis. Differences in the
efficiency with respect to unity are included in
a systematic uncertainty. The width of a 3-jet mass bin is based on the 3-jet
mass resolution, which is derived from a detector simulation.
Starting at $\mjjj = 50\GeV$, the bin width increases progressively
with \mjjj. In addition, the phase space is split into an inner,
$\aymax < 1$, and an outer, $1 \leq \aymax < 2$, rapidity region. The
bin widths in \ymax are equal to 2. Events with $\aymax \geq 2$ are
rejected.

To remove the impact of detector effects from limited acceptance and
finite resolution, the measurement is corrected with the iterative
d'Agostini unfolding algorithm~\cite{D'Agostini:1994zf} with four
iterations. Response matrices for the unfolding algorithm are derived
from detector simulation by using the two event generators \PYTHIA
version~6.4.22~\cite{Sjostrand:2006za} with tune
Z2~\cite{Field:2010bc} and \HERWIGpp
version~2.4.2~\cite{Bahr:2008pv} with the default tune. (The \PYTHIA6 Z2 tune is identical to the Z1 tune described in~\cite{Field:2010bc} except that Z2 uses the CTEQ6L PDF while Z1 uses CTEQ5L.)
Differences in the unfolding result are used to evaluate the uncertainties related
to assumptions in modelling the parton showering~\cite{Sjostrand:2004ef,
  Gieseke:2003rz}, hadronization~\cite{Andersson:1983ia,
  Andersson:1983jt, Sjostrand:1984iu, Webber:1983if}, and the
underlying event~\cite{Sjostrand:1987su, Sjostrand:2004ef,
  Bahr:2008dy} in these event generators. Additional uncertainties are
determined from an ensemble of Monte Carlo (MC) experiments, where the
data input and the response matrix are varied within the limits of
their statistical precision before entering the unfolding
algorithm. The unfolding result corresponds to the sample mean, while
the statistical uncertainty, which is propagated through the unfolding
procedure, is given by the sample covariance. The variation of the
input data leads to the statistical uncertainty in the unfolded cross
section, while the variation of the response matrix is an additional
uncertainty inherent in the unfolding technique because of the limited
size of simulated samples.

The systematic uncertainty related to the determination of the jet
energy scale (JES) is evaluated via 16 independent sources as
described in Ref.~\cite{CMS-PAPERS-QCD-11-004}. The modified prescription
for the treatment of correlations as recommended in
Ref.~\cite{CMS-PAPER-SMP-12-028} is applied.
To reduce artifacts caused by trigger turn-ons and prescale weights,
the JES uncertainty is propagated to the cross section
measurement by employing an ensemble of MC experiments, where the data
input is varied within the limits of the systematic uncertainty and
where average prescale weights are used.

The luminosity uncertainty, which is fully correlated across all \mjjj
and \ymax bins, is estimated to be 2.2\%~\cite{CMS-PAS-SMP-12-008}.

Residual jet reconstruction and trigger inefficiencies are accounted
for by an additional uncorrelated uncertainty of 1\% as in
Ref.~\cite{CMS-PAPERS-QCD-11-004}.

\begin{figure}[tbp]
  \includegraphics[width=0.48\textwidth]{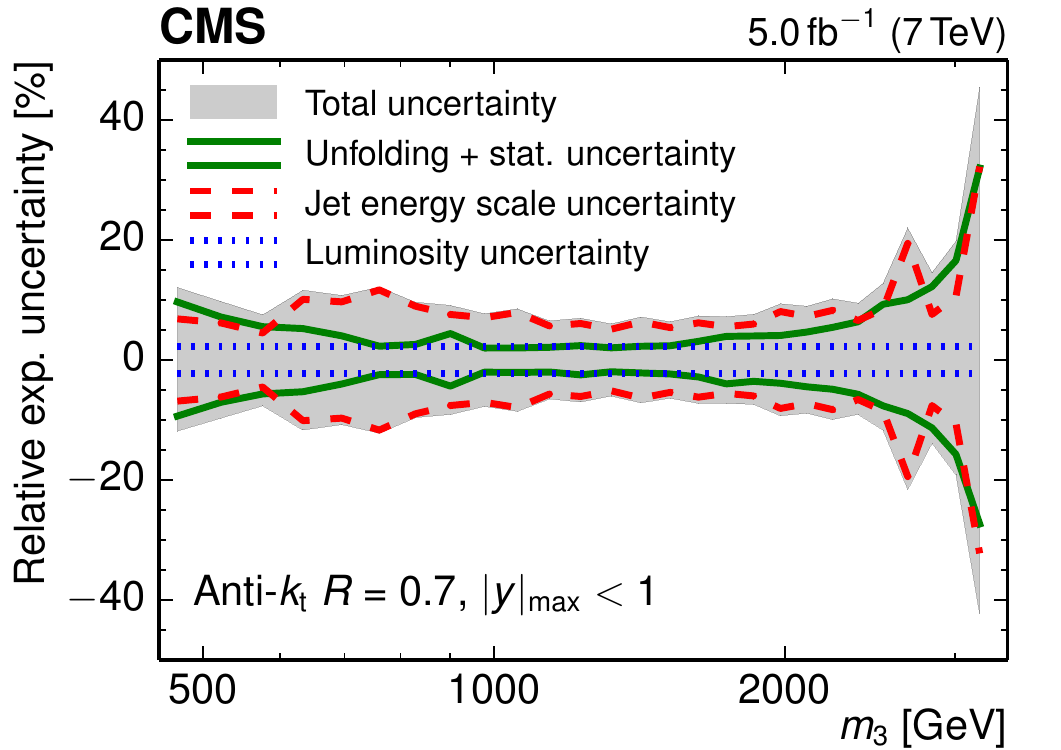}
  \includegraphics[width=0.48\textwidth]{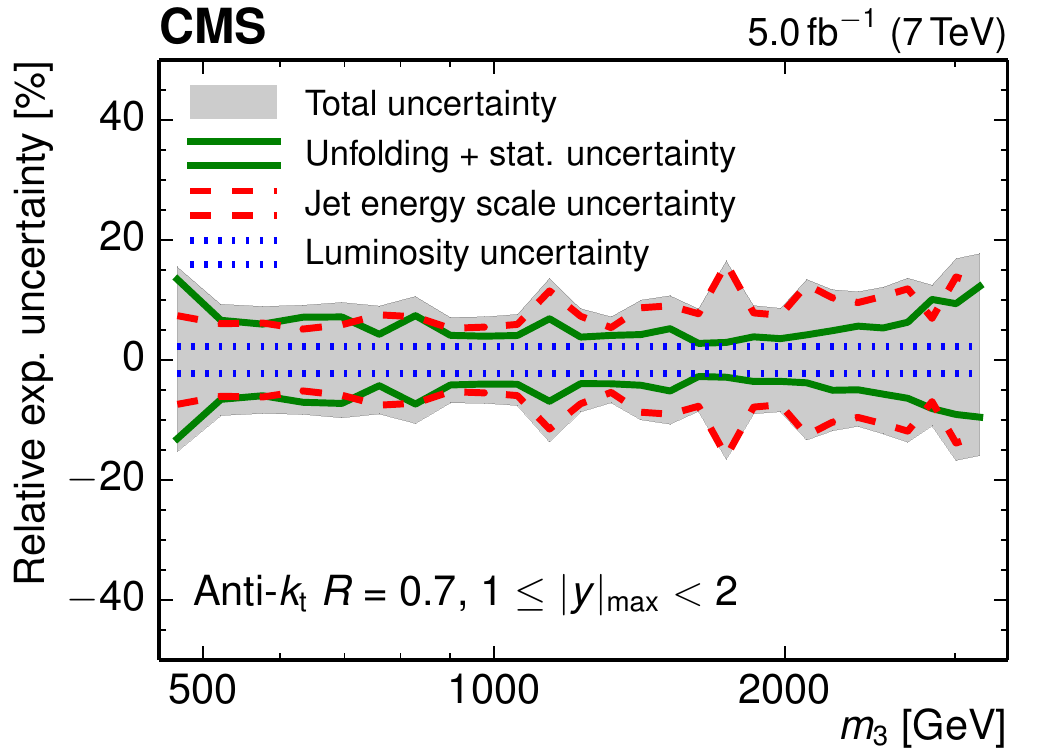}
  \caption{Overview of the measurement uncertainties in the inner
    $\aymax < 1$ (\cmsLeft) and the outer rapidity region $1 \leq \aymax <
    2$ (\cmsRight). All uncertainty components, including the 1\%
    uncorrelated residual uncertainty, are added in quadrature to give
    the total uncertainty.}
  \label{fig:exp_unc}
\end{figure}

Figure~\ref{fig:exp_unc} presents an overview of the experimental
uncertainties for the 3-jet mass measurement. Over a wide range of
3-jet masses, the JES uncertainty represents the largest
contribution. At the edges of the investigated phase space, \ie in
the low and high 3-jet mass regions, statistical and unfolding
uncertainties, which are intrinsically linked through the unfolding
procedure, become major contributors to the total uncertainty.

\section{Theoretical predictions and uncertainties}
\label{sec:theory}

The theoretical predictions for the 3-jet mass cross sections consist
of an NLO QCD calculation and a nonperturbative (NP) correction to
account for the underlying event modelled by multiparton interactions
(MPI) and for hadronization effects. Electroweak corrections to
inclusive and dijet cross sections have been calculated in
Ref.~\cite{Dittmaier:2012kx}, where they are found to be limited to
a few percent at the highest dijet masses accessible with the CMS data
at 7\TeV centre-of-mass energy. For 3-jet quantities these corrections
are not known and hence cannot be considered in the present analysis.

The NLO calculations are performed by using the \NLOJETPP program
version~4.1.3~\cite{Nagy:2001fj,Nagy:2003tz} within the framework of
the \fastNLO package version~2.1~\cite{Britzger:2012bs}. The partonic
events are subjected to the same jet algorithm and phase space
selections as the data events, where at least three jets with $\abs{y}\leq 3$
and $\pt > 100\GeV$ are required.
The number of massless quark flavours, $N_f$, is set to five. The
impact of jet production via massive top-antitop quark pairs is
estimated to be negligible. The renormalization and factorization
scales, \mur and \muf, are identified with $\mjjj/2$. With this
choice, which is identical to the jet \pt in case of dijet events at
central rapidity with $m_2/2$ as scale, the NLO corrections to the LO
cross sections remain limited between 1.2 and 1.6. The uncertainty in
the predicted cross section associated with the renormalization and
factorization scale choice is evaluated by varying \mur and \muf from
the default by the following six combinations:
$(\mur/(\mjjj/2),\muf/(\mjjj/2)) = (1/2,1/2)$, $(1/2,1)$, $(1,1/2)$,
$(1,2)$, $(2,1)$, and $(2,2)$.

Comparisons to the NLO predictions are performed for five different
PDF sets, each with NLO and NNLO PDF evolutions, from the
LHAPDF package~\cite{Whalley:2005nh}. They are listed in
Table~\ref{tab:pdfsets} together with the corresponding number of
active flavours, $N_f$, the default values of the strong coupling
constant \alpsmz, and the ranges in \alpsmz available for fits. All
PDF sets include a maximum of five active flavours $N_f$ except for
NNPDF2.1, which has $N_{f,\text{max}} = 6$. Only the ABM11 PDF set
employs a fixed-flavour number scheme in contrast to variable-flavour
number schemes favoured by all other PDF sets. The PDF uncertainties
in the cross section predictions are evaluated according to the
prescriptions recommended for the respective PDFs. More details are
available in the references listed in Table~\ref{tab:pdfsets}.

\begin{table*}[tbp]
  \caption{The PDF sets used in comparisons to the data together with the
    evolution order (Evol.), the corresponding number of active flavours,
    $N_f$, the assumed masses $M_\cPqt$ and $M_\cPZ$ of the top
    quark and the \cPZ\ boson, respectively, the default values of
    \alpsmz, and the range in \alpsmz variation available for fits. For
    CT10 the updated versions of 2012 are taken.}
  \label{tab:pdfsets}
  \centering
  \begin{tabular}{lllrclcc}
    Base set & Refs. & Evol.\ & \multicolumn{1}{r}{$N_f$} & $M_\cPqt$ [\GeVns] &
    $M_\cPZ$ [\GeVns] &\alpsmz & \alpsmz range\rbthm\\
    \hline
    ABM11     & \cite{Alekhin:2012ig} & NLO  &       5  & 180 & 91.174 & 0.1180 & 0.110--0.130\rbtrr\\
    ABM11     & \cite{Alekhin:2012ig} & NNLO &       5  & 180 & 91.174 & 0.1134 & 0.104--0.120\rbtrr\\
    CT10      & \cite{Lai:2010vv}     & NLO  & $\leq$5 & 172 & 91.188 & 0.1180 & 0.112--0.127\rbtrr\\
    CT10      & \cite{Lai:2010vv}     & NNLO & $\leq$5 & 172 & 91.188 & 0.1180 & 0.110--0.130\rbtrr\\
    HERAPDF1.5& \cite{Aaron:2009aa}   & NLO  & $\leq$5 & 180 & 91.187 & 0.1176 & 0.114--0.122\rbtrr\\
    HERAPDF1.5& \cite{Aaron:2009aa}   & NNLO & $\leq$5 & 180 & 91.187 & 0.1176 & 0.114--0.122\rbtrr\\
    MSTW2008  & \cite{Martin:2009iq,Martin:2009bu} & NLO  & $\leq$5 & $10^{10}$ & 91.1876 & 0.1202 & 0.110--0.130\rbtrr\\
    MSTW2008  & \cite{Martin:2009iq,Martin:2009bu} & NNLO & $\leq$5 & $10^{10}$ & 91.1876 & 0.1171 & 0.107--0.127\rbtrr\\
    NNPDF2.1  & \cite{Ball:2011mu}    & NLO  & $\leq$6 & 175 & 91.2 & 0.1190 & 0.114--0.124\rbtrr\\
    NNPDF2.1  & \cite{Ball:2011mu}    & NNLO & $\leq$6 & 175 & 91.2 & 0.1190 & 0.114--0.124\rbtrr\\
  \end{tabular}
\end{table*}

\begin{figure}[tbp]
  \centering
  \includegraphics[width=0.48\textwidth]{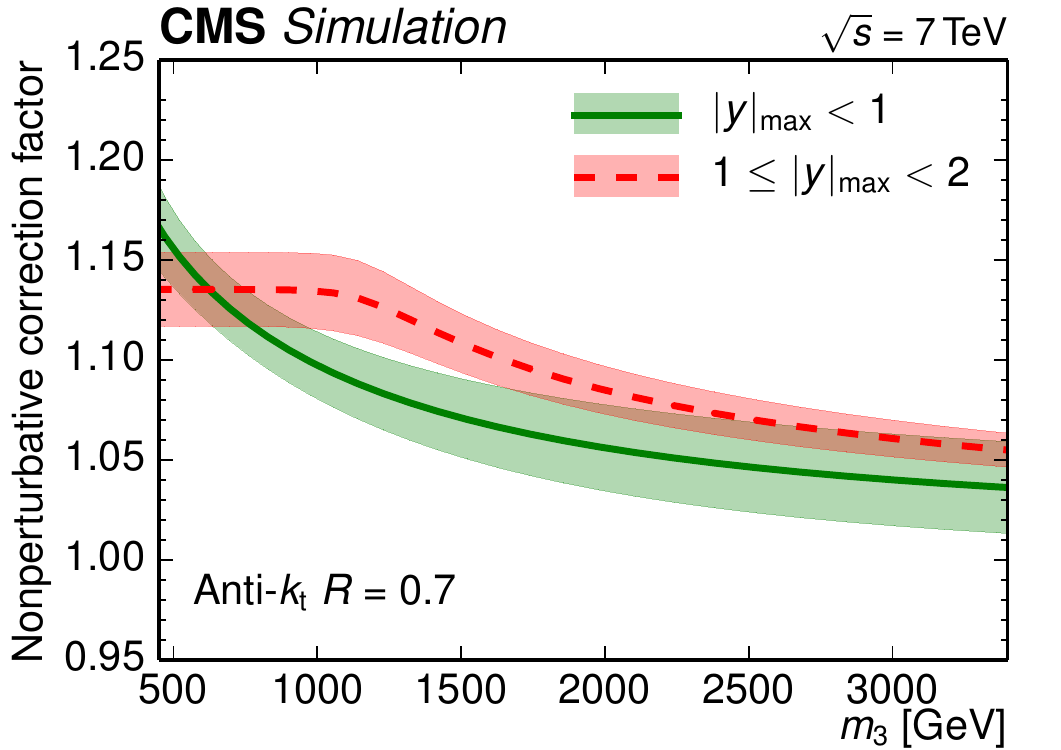}
  \caption{Overview of the NP correction factors and their
    uncertainties in the inner $\aymax < 1$ (solid line) and in the
    outer rapidity region $1 \leq \aymax < 2$ (dashed line).}
  \label{fig:np}
\end{figure}

For the NP corrections, the multijet-improved MC event generators
\SHERPA version~1.4.3~\cite{Gleisberg:2008ta} and \MADGRAPH~5 version
1.5.12~\cite{Alwall:2011uj} are used to simulate 3-jet events.
\SHERPA employs a dipole formulation for parton
showering~\cite{Winter:2007ye, Schumann:2007mg}, a cluster model for
hadronization~\cite{Winter:2003tt}, and an MPI model for the
underlying event that is based on independent hard processes similar
to \PYTHIA~\cite{Sjostrand:1987su, Gleisberg:2008ta}. In the case
of \MADGRAPH, the steps of parton showering, hadronization, and
multiple parton scatterings come from \PYTHIA version~6.4.26
with default settings using the Lund string model for
hadronization~\cite{Andersson:1983ia, Andersson:1983jt,
  Sjostrand:1984iu} and a multiple-interaction model for the
underlying event that is interleaved with the parton
shower~\cite{Sjostrand:2004ef}. The 3-jet mass is determined for a
given event before and after the MPI and hadronization phases are
performed. This allows the derivation of correction factors, which are
applied to the theory prediction at NLO\@. The correction factor is
defined as the mean of the corrections from the two examined event
generators and ranges in value from 1.16 for the low mass range to
about 1.05 at high 3-jet mass. The systematic uncertainty in the NP
correction factors is estimated as plus or minus half of the spread between
the two predictions and amounts to roughly $\pm$2\%. The NP
correction factors and their uncertainties are shown in
Fig.~\ref{fig:np} for both rapidity bins.

\begin{figure}[tbp]
  \includegraphics[width=0.48\textwidth]{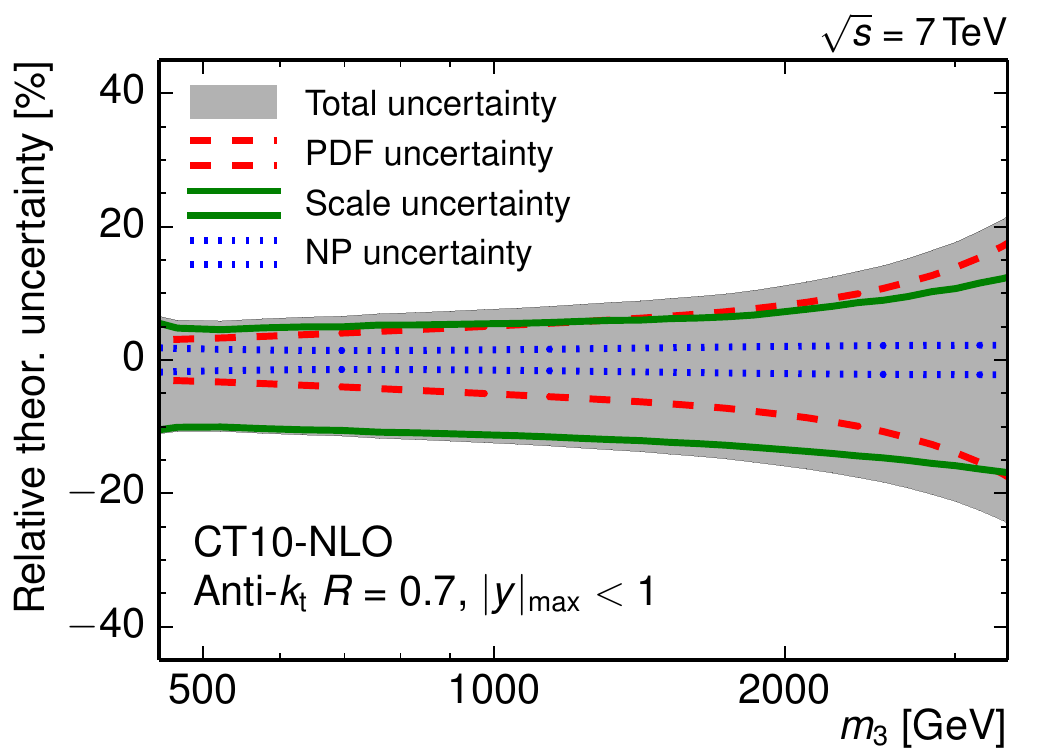}
  \includegraphics[width=0.48\textwidth]{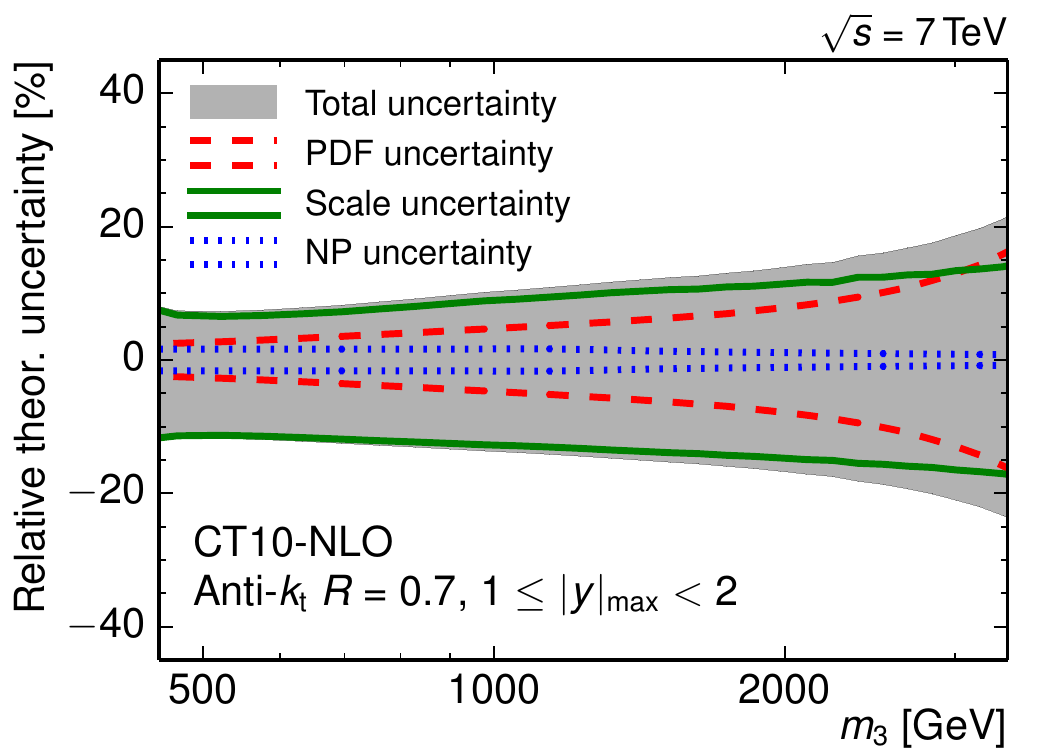}
  \caption{Overview of the theory uncertainties in the inner $\aymax <
    1$ (\cmsLeft) and in the outer rapidity region $1 \leq \aymax < 2$
    (\cmsRight) for the CT10 PDF set with NLO PDF evolution.}
  \label{fig:unc_th}
\end{figure}

An overview of the different theoretical uncertainties is given in
Fig.~\ref{fig:unc_th}.

\section{Results and determination of the strong coupling constant}
\label{sec:results}

\begin{figure}[tbp]
\centering
  \includegraphics[width=\cmsFigWidth]{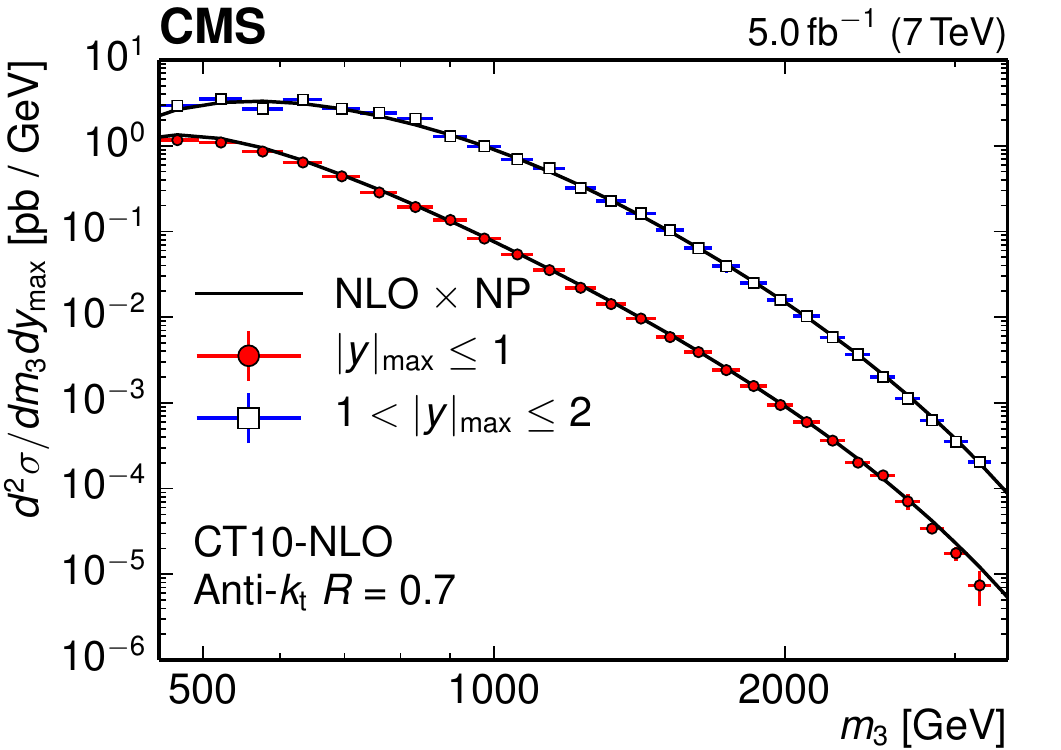}
  \caption{Comparison of the measured 3-jet mass cross section with
    the theory prediction for the two regions in \aymax.
    This prediction is based on an NLO 3-jet calculation,
    which employs the CT10-NLO PDF set and is corrected
    for nonperturbative effects. The vertical
    error bars represent the total experimental uncertainty, while the
    horizontal error bars indicate the bin widths.}
  \label{fig:comp_nlo_data}
\end{figure}

\begin{figure*}[tbp]
  \includegraphics[width=0.48\textwidth]{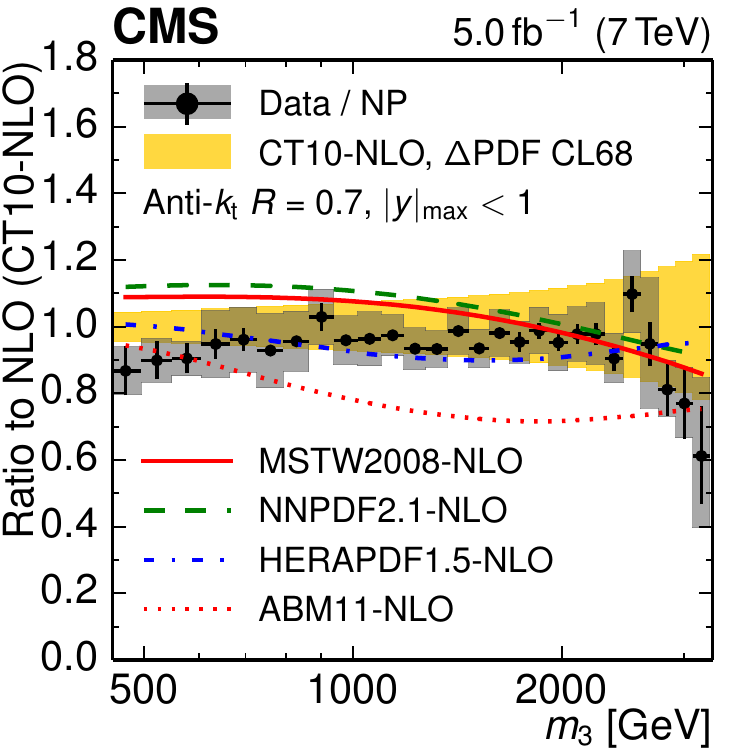}%
  \includegraphics[width=0.48\textwidth]{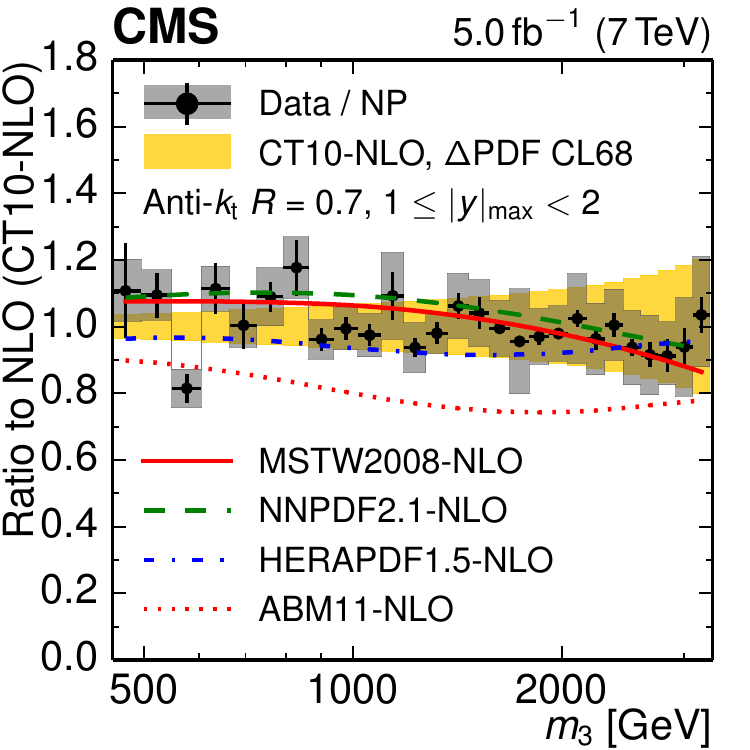}\\
  \includegraphics[width=0.48\textwidth]{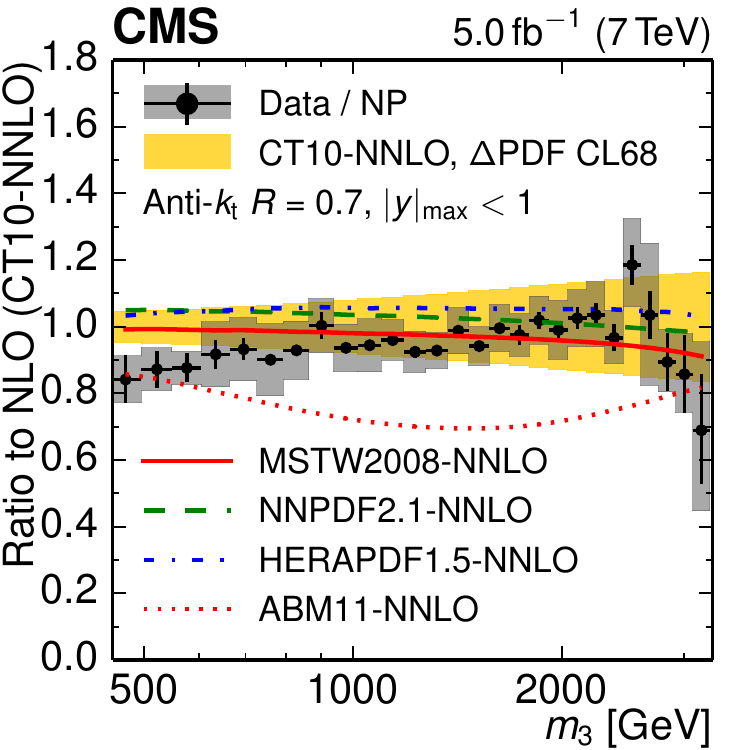}%
  \includegraphics[width=0.48\textwidth]{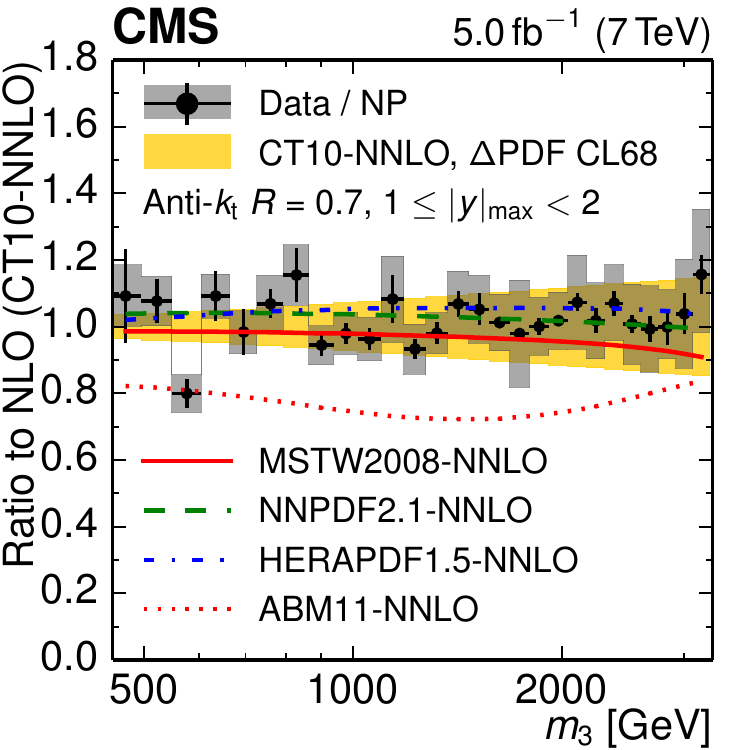}
  \caption{Ratio of the 3-jet mass cross section, divided by NP
    corrections, to the theory prediction at NLO with the CT10-NLO
    (top) or CT10-NNLO PDF set (bottom) for the inner rapidity region
    (left) and for the outer rapidity region (right).
    The data are shown with error bars
    representing the statistical uncertainty after unfolding added
    quadratically to the 1\% uncorrelated residual uncertainty and
    gray rectangles for the total correlated systematic
    uncertainty. The light gray (colour version: yellow) band
    indicates the PDF uncertainty for the CT10 PDF sets at 68\%
    confidence level. In addition, the ratios of the NLO predictions
    are displayed for the PDF sets MSTW2008, NNPDF2.1, HERAPDF1.5, and
    ABM11, also at next-to- (top) and next-to-next-to-leading
    evolution order (bottom).}
  \label{fig:comp_nlo_data_ratio}
\end{figure*}

Figure~\ref{fig:comp_nlo_data} compares the measured 3-jet mass
spectrum to the Theory prediction. This prediction is based on
an NLO 3-jet calculation, which employs the CT10-NLO PDF set
and is corrected for nonperturbative effects.
Perturbative QCD describes
the 3-jet mass cross section over five orders of magnitude for
3-jet masses up to 3\TeV. The ratios of the measured cross sections to
the theory predictions are presented in
Fig.~\ref{fig:comp_nlo_data_ratio} to better judge potential
differences between data and theory. Within uncertainties, most PDF
sets are able to describe the data. Some deviations are visible for
small \mjjj. Significant deviations are exhibited when using the
ABM11 PDFs, which therefore are not considered in our fits of \alpsmz.

\begin{figure}[tbp]
  \includegraphics[width=0.48\textwidth]{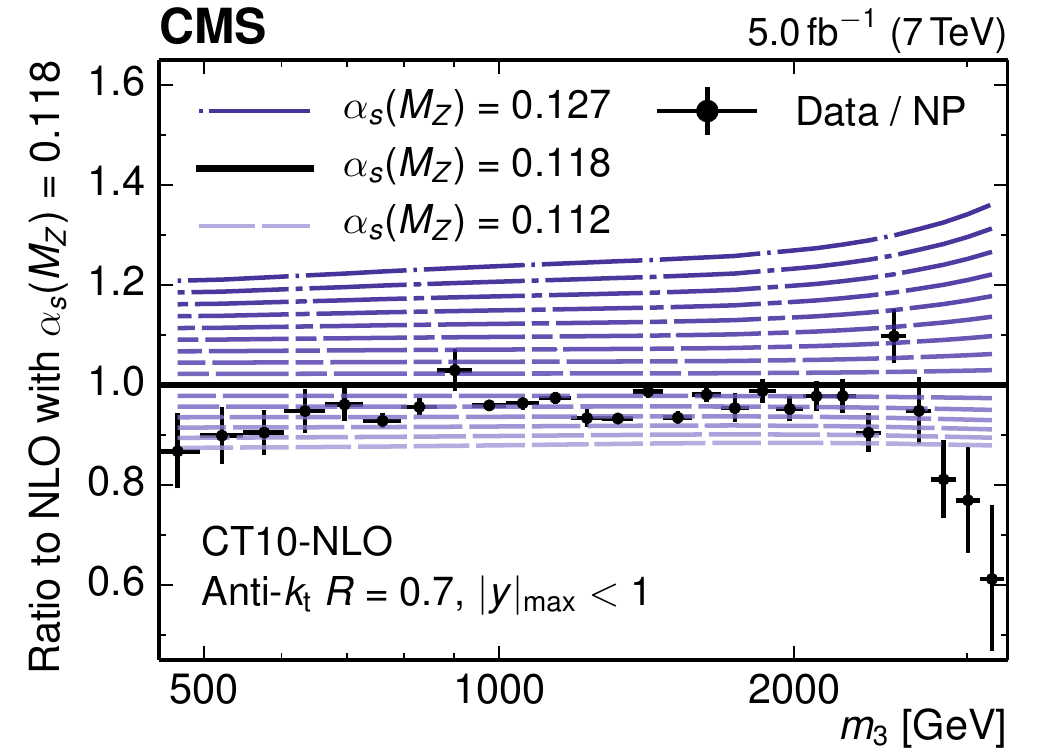}
  \includegraphics[width=0.48\textwidth]{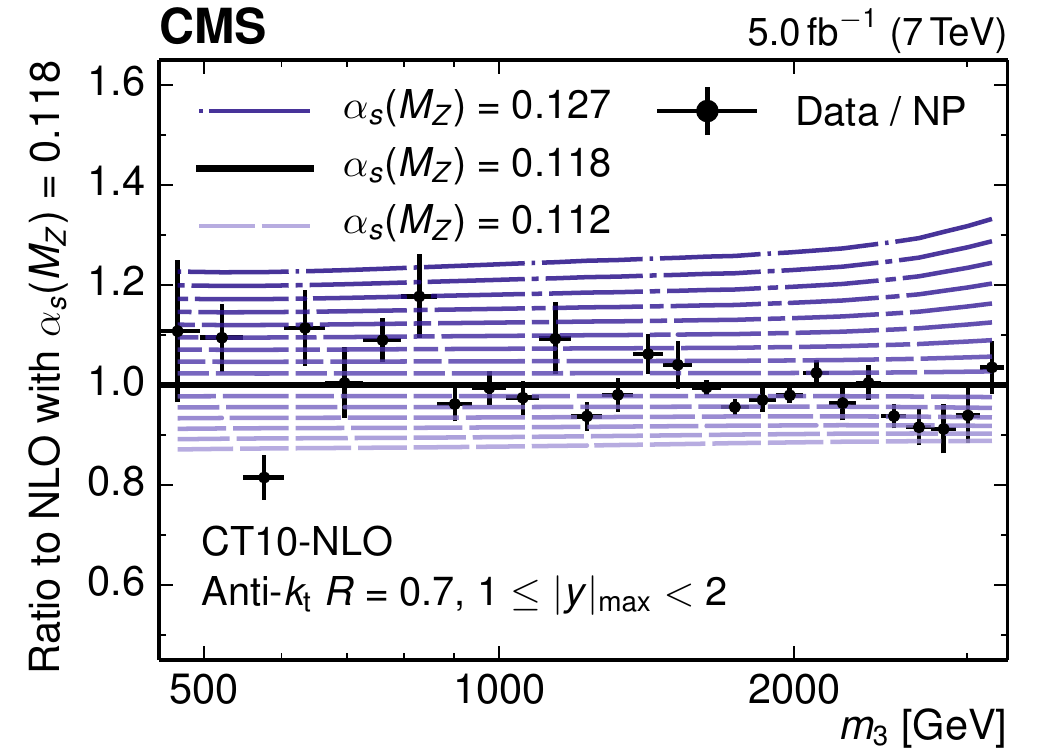}
  \caption{Ratio of the measured 3-jet mass cross section in the inner
    rapidity region (\cmsLeft) and in the outer rapidity region (\cmsRight), divided by the NP
    correction, with respect to the theory prediction at NLO while
    using the CT10-NLO PDF set with the default value of $\alpsmz =
    0.118$. In addition, ratios are shown for the theory predictions with
    CT10-NLO PDFs assuming values of \alpsmz ranging from 0.112 up to 0.127
    in steps of 0.001. The error bars
    represent the total uncorrelated uncertainty of the data.}
  \label{fig:sensitivity}
\end{figure}

In the following, the PDFs are considered to be an external input such
that a value of \alpsmz can be determined. Potential correlations
between \alpsmz and the PDFs are taken into account by using PDF sets
that include variations in \alpsmz as listed in
Table~\ref{tab:pdfsets}. Figure~\ref{fig:sensitivity} demonstrates for
the example of the CT10-NLO PDF set the sensitivity of the theory
predictions with respect to variations in the value of \alpsmz in
comparison to the data and their total uncertainty.

A value of \alpsmz is determined by minimizing the \chisq between the
$N$ measurements $D_i$ and the theoretical predictions $T_i$. The
\chisq is defined as
\begin{equation}
  \chisq = \sum_{ij}^N \left(D_i - T_i\right) \mathrm{C}_{ij}^{-1}
  \left(D_j - T_j\right),
  \label{chi2_square}
\end{equation}
where the covariance matrix $C_{ij}$ is composed of the following
terms:
\ifthenelse{\boolean{cms@external}}{
\begin{multline}
  C = \cov_{\text{unf}+\text{stat}} +
  \cov_{\text{uncor}} +\\
  \left(\sum_\text{sources}\cov_\mathrm{JES}\right) +
  \cov_{\text{lumi}} +
  \cov_{\mathrm{PDF}},
\end{multline}
}{
\begin{equation}
  C = \cov_{\text{unf}+\text{stat}} +
  \cov_{\text{uncor}} +
  \left(\sum_\text{sources}\cov_\mathrm{JES}\right) +
  \cov_{\text{lumi}} +
  \cov_{\mathrm{PDF}},
\end{equation}
}
and the terms in the sum represent
\begin{enumerate}
\item{$\cov_{\text{unf}+\text{stat}}$: statistical and
    unfolding uncertainty including correlations induced through the
    unfolding};
\item{$\cov_\text{uncor}$: uncorrelated systematic
    uncertainty summing up small residual effects such as trigger and
    identification inefficiencies, time dependence of the jet \pt
    resolution, and the uncertainty on the trigger prescale factor};
\item{$\cov_{\mathrm{JES},\text{sources}}$: systematic uncertainty for
    each JES uncertainty source};
\item{$\cov_\text{lumi}$: luminosity uncertainty}; and
\item{$\cov_\mathrm{PDF}$: PDF uncertainties}.
\end{enumerate}

The first four sources constitute the experimental uncertainty.
The JES and luminosity uncertainty are treated as fully correlated
across the \mjjj and \aymax bins, where for the JES uncertainty the
procedure recommended in Ref.~\cite{CMS-PAPER-SMP-12-028} is applied.
The derivation of PDF uncertainties follows prescriptions for each
individual PDF set. The CT10 and MSTW PDF sets both employ the Hessian
or eigenvector method~\cite{Pumplin:2001ct} with upward and downward
variations for each eigenvector. As required by the use of covariance
matrices, symmetric PDF uncertainties are computed following
Ref.~\cite{Pumplin:2002vw}. For the HERAPDF1.5 PDF set, which employs a
Hessian method for the experimental uncertainties,
complemented with model and parameterization uncertainties, the
prescription from Ref.~\cite{Aaron:2009aa} is used. The NNPDF2.1 PDF
set uses the technique of MC pseudo-experiments instead of the
eigenvector method to provide PDF uncertainties. The ensemble of
replicas, whose averaged predictions give the central result, are
evaluated following the prescription in Ref.~\cite{Ball:2010de} to
derive the PDF uncertainty for NNPDF\@. The JES and luminosity
uncertainties are assumed to be multiplicative to avoid the
statistical bias that arises from uncertainty estimations taken from
data~\cite{Lyons:1989gh,D'Agostini:2003nk,Ball:2009qv}. The
uncertainty in a result for \alpsmz from a \chisq fit is obtained
from the \alpsmz values for which the \chisq is increased by one with
respect to the minimum value.

The uncertainty in \alpsmz due to the NP uncertainties is evaluated by
looking for maximal offsets from a default fit. The theoretical
prediction $T$ is varied by the NP uncertainty $\Delta\mathrm{NP}$ as
$T\cdot\mathrm{NP} \to T\cdot\left(\mathrm{NP} \pm
  \Delta\mathrm{NP}\right)$. The fitting procedure is repeated for
these two variations, and the deviation from the central \alpsmz
values is considered as the uncertainty in \alpsmz. Finally, the
uncertainty due to the \mur and \muf scales is evaluated by applying
the same method as for the NP corrections, varying \mur and \muf by the
six scale factor combinations as described in
Section~\ref{sec:theory}.

The shape of the predicted 3-jet mass cross section depends on the QCD
matrix elements and kinematic constraints. Because each of the leading
three jets is required to have a \pt larger than 100\GeV, some event
configurations, possible with respect to the QCD matrix elements, are
kinematically forbidden at low \mjjj. In the spectra shown in
Fig.~\ref{fig:comp_nlo_data}, this fact is visible in the form of a
maximum in the 3-jet mass cross section, which is shifted to higher
\mjjj values for the outer compared to the inner \aymax bin because
the larger differences in the jet rapidities allow higher \mjjj to be
reached with lower \pt jets. For fits of \alpsmz the \mjjj region
limited through kinematical constraints is unsuited, since close to the
phase space boundaries fixed-order pQCD calculations might be
insufficient and resummations might be required. To avoid this region
of phase space as done in Ref.~\cite{CMS-PAPERS-QCD-11-003}, only
\mjjj bins beyond the maximum of the 3-jet mass cross section in the
outer \aymax bin are considered. This corresponds to a minimum in
\mjjj of 664\GeV. Including one bin more or less induces changes only
in the measured \alpsmz below the percent level. To study the running of the strong coupling, the
comparison between data and theory is also performed in several 3-jet
mass regions above 664\GeV as shown in Table~\ref{tab:asfits}.

For the evolution of \alpsq in the fits of \alpsmz, the Gl{\"u}ck--Reya--Vogt
formula~\cite{Gluck:1998xa} is used at 2-loop order as implemented in
\fastNLO. The capability of \fastNLO to replace the \alpsq evolution
of a PDF set by such alternative codes is exploited to interpolate
cross section predictions between the available fixed points of
\alpsmz listed in Table~\ref{tab:pdfsets}.
Limited extrapolations beyond the lowest or highest values
of \alpsmz provided in a PDF series are accepted if necessary for
uncertainty evaluations, up to a limit of $|\Delta\alpsmz| =
0.003$. This extrapolation method can be necessary in some cases to fully evaluate
the scale uncertainty. The procedure has been cross-checked using the
original \alpsq grid of each PDF within LHAPDF and with the evolution
code of the \HOPPET toolkit~\cite{Salam:2008qg} and of
\RunDec~\cite{Schmidt:2012az,Chetyrkin:2000yt}.

The CT10-NLO PDF set is chosen for the main result for two reasons:
The range in available \alpsmz values is wide enough to evaluate
almost all scale uncertainties within this range and the central value
of \alpsmz in this set is rather close to the combined fit result.

The fit results for \alpsmz and \alpsq for all considered \mjjj ranges
are presented in Tables~\ref{tab:asfits} and~\ref{tab:asqfits},
respectively. Fits
over the total \mjjj range above 664\GeV are shown for each \ymax bin
separately and for both combined in the bottom three rows of Table~\ref{tab:asfits}.

\begin{table*}[tbp]
  \topcaption{Determinations of \alpsmz in the considered \mjjj ranges. The
    relevant scale in each 3-jet mass range is calculated from the
    cross section-weighted average as given by the theory prediction using
    the CT10 PDF set with NLO evolution. The three bottom rows present
    fits using the whole 3-jet mass range above 664\GeV in both
    rapidity regions either separately or combined (last
    row). Uncertainties are quoted separately for experimental sources,
    the PDFs, the NP corrections, and the scale uncertainty.}
  \label{tab:asfits}
  \centering
  \begin{tabular}{rrrccccc}
    \mjjj [{\GeVns}] &
    \multicolumn{1}{c}{$\left<Q\right>$ [{\GeVns}]} &
    \multicolumn{1}{c}{$\chisqndof$} &
    $\alpsmz$ & $\pm \text{(exp)}$ & $\pm \mathrm{(PDF)}$ & $\pm \mathrm{(NP)}$ & $\pm\text{(scale)}$ \rbthm\\\hline
    $ 664$--$ 794$ & $ 361 $ & $ 4.5 /   3$ & $0.1232$ & $^{+0.0040}_{-0.0042} $ & $^{+0.0019}_{-0.0016} $ & $^{+0.0008}_{-0.0007} $ & $^{+0.0079}_{-0.0044} $ \rbtrr\\
    $ 794$--$ 938$ & $ 429 $ & $ 7.8 /   3$ & $0.1143$ & $^{+0.0034}_{-0.0033} $ & $^{+0.0019}_{-0.0016} $ & $\pm 0.0008            $ & $^{+0.0073}_{-0.0042} $ \rbtrr\\
    $ 938$--$1098$ & $ 504 $ & $ 0.6 /   3$ & $0.1171$ & $^{+0.0033}_{-0.0034} $ & $\pm 0.0022            $ & $\pm 0.0007            $ & $^{+0.0068}_{-0.0040} $ \rbtrr\\
    $1098$--$1369$ & $ 602 $ & $ 2.6 /   5$ & $0.1152$ & $\pm 0.0026            $ & $^{+0.0027}_{-0.0026} $ & $^{+0.0008}_{-0.0007} $ & $^{+0.0060}_{-0.0027} $ \rbtrr\\
    $1369$--$2172$ & $ 785 $ & $ 8.8 /  13$ & $0.1168$ & $^{+0.0018}_{-0.0019} $ & $^{+0.0030}_{-0.0031} $ & $^{+0.0007}_{-0.0006} $ & $^{+0.0068}_{-0.0034} $ \rbtrr\\
    $2172$--$2602$ & $1164 $ & $ 3.6 /   5$ & $0.1167$ & $^{+0.0037}_{-0.0044} $ & $^{+0.0040}_{-0.0044} $ & $\pm 0.0008            $ & $^{+0.0065}_{-0.0041} $ \rbtrr\\
    $2602$--$3270$ & $1402 $ & $ 5.5 /   7$ & $0.1120$ & $^{+0.0043}_{-0.0041} $ & $^{+0.0056}_{-0.0040} $ & $\pm 0.0001            $ & $^{+0.0088}_{-0.0050} $ \rbtrr\\
    \hline
    $\aymax < 1$   & $ 413 $ & $10.3 /  22$ & $0.1163$ & $^{+0.0018}_{-0.0019} $ & $\pm 0.0027            $ & $\pm 0.0007            $ & $^{+0.0059}_{-0.0025} $ \rbtrr\\
    $1\leq\aymax<2$& $ 441 $ & $10.6 /  22$ & $0.1179$ & $^{+0.0018}_{-0.0019} $ & $\pm 0.0021            $ & $\pm 0.0007            $ & $^{+0.0067}_{-0.0037} $ \rbtrr\\
    $\aymax < 2$   & $ 438 $ & $47.2 /  45$ & $0.1171$ & $\pm 0.0013            $ & $\pm 0.0024            $ & $\pm 0.0008            $ & $^{+0.0069}_{-0.0040} $ \rbtrr\\
  \end{tabular}
\end{table*}

\begin{table*}[tbp]
  \topcaption{Same as Table~\ref{tab:asfits} but showing the fit result
    in terms of \alpsq for each range in $Q$.}
  \label{tab:asqfits}
  \centering
  \begin{tabular}{rrrccccc}
    \mjjj [\GeVns{}] &
    \multicolumn{1}{c}{$\left<Q\right>$ [\GeVns{}]} &
    \multicolumn{1}{c}{$\chisqndof$} &
    $\alpsq$ & $\pm \text{(exp)}$ & $\pm \mathrm{(PDF)}$ & $\pm \mathrm{(NP)}$ & $\pm\text{(scale)}$ \rbthm\\\hline
    $ 664$--$ 794$ & $ 361 $ & $ 4.5 /   3$ & $0.1013$ & $^{+0.0027}_{-0.0028} $ & $^{+0.0013}_{-0.0011} $ & $\pm 0.0005            $ & $^{+0.0052}_{-0.0030} $ \rbtrr\\
    $ 794$--$ 938$ & $ 429 $ & $ 7.8 /   3$ & $0.0933$ & $\pm 0.0022            $ & $^{+0.0012}_{-0.0011} $ & $\pm 0.0005            $ & $^{+0.0048}_{-0.0028} $ \rbtrr\\
    $ 938$--$1098$ & $ 504 $ & $ 0.6 /   3$ & $0.0934$ & $\pm 0.0021            $ & $\pm 0.0014            $ & $\pm 0.0005            $ & $^{+0.0043}_{-0.0025} $ \rbtrr\\
    $1098$--$1369$ & $ 602 $ & $ 2.6 /   5$ & $0.0902$ & $\pm 0.0016            $ & $\pm 0.0016            $ & $^{+0.0005}_{-0.0004} $ & $^{+0.0036}_{-0.0017} $ \rbtrr\\
    $1369$--$2172$ & $ 785 $ & $ 8.8 /  13$ & $0.0885$ & $^{+0.0010}_{-0.0011} $ & $^{+0.0017}_{-0.0018} $ & $^{+0.0004}_{-0.0003} $ & $^{+0.0038}_{-0.0020} $ \rbtrr\\
    $2172$--$2602$ & $1164 $ & $ 3.6 /   5$ & $0.0848$ & $^{+0.0019}_{-0.0023} $ & $^{+0.0020}_{-0.0023} $ & $\pm 0.0004            $ & $^{+0.0034}_{-0.0021} $ \rbtrr\\
    $2602$--$3270$ & $1402 $ & $ 5.5 /   7$ & $0.0807$ & $^{+0.0022}_{-0.0021} $ & $^{+0.0028}_{-0.0021} $ & $\pm 0.0001            $ & $^{+0.0044}_{-0.0026} $ \rbtrr\\
  \end{tabular}
\end{table*}

For comparison, the combined fit was also tried for alternative PDF
sets listed in Table~\ref{tab:asfits_pdf}. For the ABM11 PDFs,
which predict 3-jet mass cross sections that are too small, fits are
technically possible. However, to compensate for this discrepancy, the
\alpsmz results take unreasonably high values that are far outside
the \alpsmz values that are given by the PDF authors. For the
NNPDF2.1-NLO and HERAPDF1.5-NLO PDF series, a central value for
\alpsmz can be calculated, but the range in \alpsmz values is not
sufficient for a reliable determination of uncertainty estimations.
In all other cases the fit results for \alpsmz are in agreement
between the investigated PDF sets and PDF evolution orders within
uncertainties.

\begin{table*}[tbp]
  \topcaption{%
    Determinations of \alpsmz with different PDF sets using all 3-jet
    mass points with $\mjjj > 664\GeV$. Uncertainties are quoted
    separately for experimental sources, the PDFs, the NP corrections,
    and the scale uncertainty.}
  \label{tab:asfits_pdf}
  \centering
  \begin{tabular}{lccccccccc}
    \multicolumn{1}{c}{PDF set} &
    $\chisqndof$ &
    $\alpsmz$ & $\pm \text{(exp)}$ & $\pm \mathrm{(PDF)}$ & $\pm \mathrm{(NP)}$ & $\pm\scale$ \rbthm\\\hline
    CT10-NLO & $47.2 /  45$ & $0.1171$ & $\pm 0.0013            $ & $\pm 0.0024            $ & $\pm 0.0008            $ & $^{+0.0069}_{-0.0040} $ \rbtrr\\
    CT10-NNLO & $48.5 /  45$ & $0.1165$ & $^{+0.0011}_{-0.0010} $ & $^{+0.0022}_{-0.0023} $ & $^{+0.0006}_{-0.0008} $ & $^{+0.0066}_{-0.0034} $ \rbtrr\\
    MSTW2008-NLO & $52.8 /  45$ & $0.1155$ & $^{+0.0014}_{-0.0013} $ & $^{+0.0014}_{-0.0015} $ & $^{+0.0008}_{-0.0009} $ & $^{+0.0105}_{-0.0029} $ \rbtrr\\
    MSTW2008-NNLO & $53.9 /  45$ & $0.1183$ & $^{+0.0011}_{-0.0016} $ & $^{+0.0012}_{-0.0023} $ & $^{+0.0011}_{-0.0019} $ & $^{+0.0052}_{-0.0050} $ \rbtrr\\
    HERAPDF1.5-NNLO& $49.9 /  45$ & $0.1143$ & $\pm 0.0007            $ & $^{+0.0020}_{-0.0035} $ & $^{+0.0003}_{-0.0008} $ & $^{+0.0035}_{-0.0027} $ \rbtrr\\
    NNPDF2.1-NNLO& $51.1 /  45$ & $0.1164$ & $\pm 0.0010            $ & $^{+0.0020}_{-0.0019} $ & $^{+0.0010}_{-0.0009} $ & $^{+0.0058}_{-0.0025} $ \rbtrr\\
  \end{tabular}
\end{table*}

\begin{figure}[tbp]
  \centering
  \includegraphics[width=\cmsFigWidth]{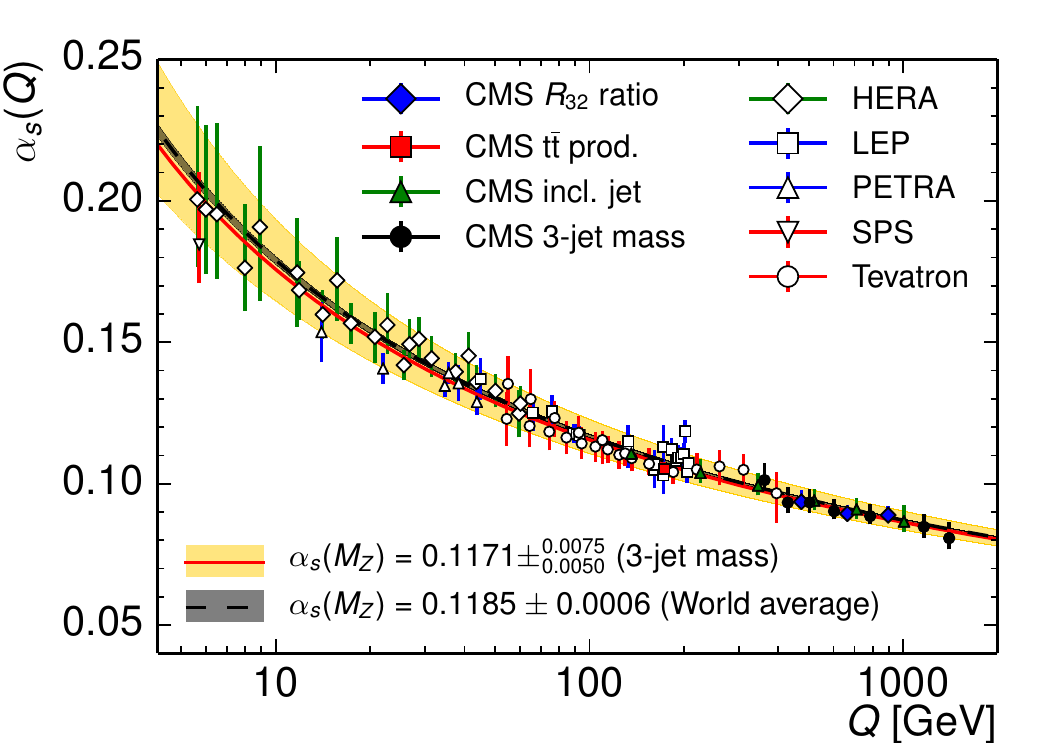}
  \caption{Comparison of the \alpsq evolution as determined in this
    analysis from all measurement bins with $\mjjj > 664\GeV$ (solid
    curve with light grey uncertainty band; colour version: red curve
    with yellow uncertainty band) to the world average (dashed curve
    with dark grey uncertainty band)~\cite{Agashe:2014kda}. The error
    bars on the data points correspond to the total uncertainty. In
    addition, an overview of measurements of the running of the strong
    coupling $\alpsq$ from electron-positron~\cite{Schieck:2006tc,Abdallah:2004xe, MartiiGarcia:1997uf},
    electron-proton~\cite{Aaron:2009vs, Aaron:2010ac,Abramowicz:2012jz, Andreev:2014wwa}, and proton--(anti)proton
    collider experiments~\cite{Abazov:2009nc, Abazov:2012lua,CMS-PAPERS-QCD-11-003, Chatrchyan:2013haa} is presented. The
    results of this analysis extend the covered range in values of the
    scale $Q$ up to ${\approx}$1.4\TeV.}
  \label{fig:asfitsall}
\end{figure}

Figure~\ref{fig:asfitsall} shows the \alpsq evolution determined in
this analysis with CT10-NLO in comparison to the world average of
$\alpsmz = 0.1185 \pm 0.0006$~\cite{Agashe:2014kda}. The figure also
shows an overview of the measurements of the running of the strong
coupling from various other experiments~\cite{Abazov:2009nc,
  Abazov:2012lua, Chekanov:2006yc, Adloff:2000qk, Schieck:2006tc,
  Abdallah:2004xe, MartiiGarcia:1997uf} together with recent
determinations by CMS~\cite{CMS-PAPERS-QCD-11-003, Chatrchyan:2013haa,
  CMS-PAPER-SMP-12-028} and from this analysis. Within uncertainties,
the new results presented here are in agreement with previous
determinations and extend the covered range in scale $Q$ up to a value
of 1.4\TeV.

\section{Summary}
\label{sec:conclusions}

The proton-proton collision data collected by the CMS experiment in
2011 at a centre-of-mass energy of 7\TeV were used to measure the
double-differential 3-jet production cross section as a function of
the invariant mass \mjjj of the three jets leading in \pt, and of
their maximum rapidity \ymax. The measurement covers a 3-jet mass
range from 445\GeV up to 3270\GeV in two bins of rapidity up to
$\abs{\ymax} = 2$. Within experimental and theoretical uncertainties, which
are of comparable size, the data are in agreement with predictions of
perturbative QCD at next-to-leading order.

The strong coupling
constant has been determined in multiple regions of 3-jet mass
for values of the scale $Q$ between 0.4 and 1.4\TeV
from a comparison between data and theory. The
results are consistent with the evolution of the strong coupling as
predicted by the renormalization group equation and extend the range
in $Q$ where this could be tested up to 1.4\TeV.
A combined fit of all data points above a 3-jet mass of
664\GeV gives the value of the strong coupling constant
$\alpsmz = 0.1171 %
\allowbreak\pm 0.0013\texp %
\allowbreak\pm 0.0024\,(\mathrm{PDF}) %
\allowbreak\pm 0.0008\,(\mathrm{NP}) %
\allowbreak\,^{+0.0069}_{-0.0040}\scale$.

This result, achieved with 3-jet production cross sections, is
consistent with determinations previously reported by CMS using the inclusive
jet cross section~\cite{CMS-PAPER-SMP-12-028} and the ratio of
inclusive 3-jet to inclusive 2-jet production cross sections~\cite{CMS-PAPERS-QCD-11-003}.
It is also consistent with a recent determination of \alpsmz by CMS
at the top production threshold using theory at NNLO~\cite{Chatrchyan:2013haa} and
with the latest world average of $\alpsmz = 0.1185 \pm 0.0006$~\cite{Agashe:2014kda}.

\begin{acknowledgments}
\hyphenation{Bundes-ministerium Forschungs-gemeinschaft Forschungs-zentren} We congratulate our colleagues in the CERN accelerator departments for the excellent performance of the LHC and thank the technical and administrative staffs at CERN and at other CMS institutes for their contributions to the success of the CMS effort. In addition, we gratefully acknowledge the computing centres and personnel of the Worldwide LHC Computing Grid for delivering so effectively the computing infrastructure essential to our analyses. Finally, we acknowledge the enduring support for the construction and operation of the LHC and the CMS detector provided by the following funding agencies: the Austrian Federal Ministry of Science, Research and Economy and the Austrian Science Fund; the Belgian Fonds de la Recherche Scientifique, and Fonds voor Wetenschappelijk Onderzoek; the Brazilian Funding Agencies (CNPq, CAPES, FAPERJ, and FAPESP); the Bulgarian Ministry of Education and Science; CERN; the Chinese Academy of Sciences, Ministry of Science and Technology, and National Natural Science Foundation of China; the Colombian Funding Agency (COLCIENCIAS); the Croatian Ministry of Science, Education and Sport, and the Croatian Science Foundation; the Research Promotion Foundation, Cyprus; the Ministry of Education and Research, Estonian Research Council via IUT23-4 and IUT23-6 and European Regional Development Fund, Estonia; the Academy of Finland, Finnish Ministry of Education and Culture, and Helsinki Institute of Physics; the Institut National de Physique Nucl\'eaire et de Physique des Particules~/~CNRS, and Commissariat \`a l'\'Energie Atomique et aux \'Energies Alternatives~/~CEA, France; the Bundesministerium f\"ur Bildung und Forschung, Deutsche Forschungsgemeinschaft, and Helmholtz-Gemeinschaft Deutscher Forschungszentren, Germany; the General Secretariat for Research and Technology, Greece; the National Scientific Research Foundation, and National Innovation Office, Hungary; the Department of Atomic Energy and the Department of Science and Technology, India; the Institute for Studies in Theoretical Physics and Mathematics, Iran; the Science Foundation, Ireland; the Istituto Nazionale di Fisica Nucleare, Italy; the Ministry of Science, ICT and Future Planning, and National Research Foundation (NRF), Republic of Korea; the Lithuanian Academy of Sciences; the Ministry of Education, and University of Malaya (Malaysia); the Mexican Funding Agencies (CINVESTAV, CONACYT, SEP, and UASLP-FAI); the Ministry of Business, Innovation and Employment, New Zealand; the Pakistan Atomic Energy Commission; the Ministry of Science and Higher Education and the National Science Centre, Poland; the Funda\c{c}\~ao para a Ci\^encia e a Tecnologia, Portugal; JINR, Dubna; the Ministry of Education and Science of the Russian Federation, the Federal Agency of Atomic Energy of the Russian Federation, Russian Academy of Sciences, and the Russian Foundation for Basic Research; the Ministry of Education, Science and Technological Development of Serbia; the Secretar\'{\i}a de Estado de Investigaci\'on, Desarrollo e Innovaci\'on and Programa Consolider-Ingenio 2010, Spain; the Swiss Funding Agencies (ETH Board, ETH Zurich, PSI, SNF, UniZH, Canton Zurich, and SER); the Ministry of Science and Technology, Taipei; the Thailand Center of Excellence in Physics, the Institute for the Promotion of Teaching Science and Technology of Thailand, Special Task Force for Activating Research and the National Science and Technology Development Agency of Thailand; the Scientific and Technical Research Council of Turkey, and Turkish Atomic Energy Authority; the National Academy of Sciences of Ukraine, and State Fund for Fundamental Researches, Ukraine; the Science and Technology Facilities Council, UK; the US Department of Energy, and the US National Science Foundation.

Individuals have received support from the Marie-Curie programme and the European Research Council and EPLANET (European Union); the Leventis Foundation; the A. P. Sloan Foundation; the Alexander von Humboldt Foundation; the Belgian Federal Science Policy Office; the Fonds pour la Formation \`a la Recherche dans l'Industrie et dans l'Agriculture (FRIA-Belgium); the Agentschap voor Innovatie door Wetenschap en Technologie (IWT-Belgium); the Ministry of Education, Youth and Sports (MEYS) of the Czech Republic; the Council of Science and Industrial Research, India; the HOMING PLUS programme of Foundation for Polish Science, cofinanced from European Union, Regional Development Fund; the Compagnia di San Paolo (Torino); the Consorzio per la Fisica (Trieste); MIUR project 20108T4XTM (Italy); the Thalis and Aristeia programmes cofinanced by EU-ESF and the Greek NSRF; and the National Priorities Research Program by Qatar National Research Fund.
\end{acknowledgments}
\bibliography{auto_generated}
\cleardoublepage \appendix\section{The CMS Collaboration \label{app:collab}}\begin{sloppypar}\hyphenpenalty=5000\widowpenalty=500\clubpenalty=5000\input{SMP-12-027-authorlist.tex}\end{sloppypar}
\end{document}

%% file: SMP-12-027-authorlist.tex
\textbf{Yerevan Physics Institute,  Yerevan,  Armenia}\\*[0pt]
V.~Khachatryan, A.M.~Sirunyan, A.~Tumasyan
\vskip\cmsinstskip
\textbf{Institut f\"{u}r Hochenergiephysik der OeAW,  Wien,  Austria}\\*[0pt]
W.~Adam, T.~Bergauer, M.~Dragicevic, J.~Er\"{o}, C.~Fabjan\cmsAuthorMark{1}, M.~Friedl, R.~Fr\"{u}hwirth\cmsAuthorMark{1}, V.M.~Ghete, C.~Hartl, N.~H\"{o}rmann, J.~Hrubec, M.~Jeitler\cmsAuthorMark{1}, W.~Kiesenhofer, V.~Kn\"{u}nz, M.~Krammer\cmsAuthorMark{1}, I.~Kr\"{a}tschmer, D.~Liko, I.~Mikulec, D.~Rabady\cmsAuthorMark{2}, B.~Rahbaran, H.~Rohringer, R.~Sch\"{o}fbeck, J.~Strauss, A.~Taurok, W.~Treberer-Treberspurg, W.~Waltenberger, C.-E.~Wulz\cmsAuthorMark{1}
% ...
\vskip\cmsinstskip
% ...
1:~~Also at Vienna University of Technology, Vienna, Austria\\
2:~~Also at CERN, European Organization for Nuclear Research, Geneva, Switzerland\\
% ...
